# The Spectra of $^2$H and $^3$He Secondary Cosmic Ray Isotopes from

# ~20-85 MeV/nuc as Measured Using the B-end HET Telescope

# on Voyager beyond the Heliopause  and a Fit to These Interstellar

# Spectra Using a Leaky Box Propagation Model

W.R. Webber[1], N. Lal[2] and B. Heikkila[2]

1.  New Mexico State University, Department of Astronomy, Las Cruces, NM  88003, USA

2.  NASA/Goddard Space Flight Center, Greenbelt, MD  20771, USA

**ABSTRACT**

We have measured the intensity and spectra of the cosmic ray secondary isotopes $^2$H and $^3$He and the primary isotopes H and $^4$He between ~20-85 MeV/nuc during a 5 year time period after Voyager 1 (V1) crossed the heliopause. The data reported here is from the B-end high energy telescope. The ratios of the intensities of the secondary to primary spectra of these nuclei at low energies are sensitive indicators for determining the amount of matter traversed at these energies by their galactic cosmic ray progenitor nuclei, after acceleration, in this case mostly $^4$He nuclei. The measurements of secondary $^3$He abundances indicate that cosmic ray $^4$He of energies between 30-100 MeV/nuc have traversed between 7-9 g/cm$^2$ of interstellar matter (90% H, 10% He) in a Leaky Box propagation model. This path length is also consistent with the production of secondary $^2$H nuclei between ~20-50 MeV/nuc, which is also produced mainly by $^4$He in a LBM. The Boron abundance, studied in separate papers, is also consistent with this path length at energies >30 MeV/nuc. These secondary intensities imply that the interstellar cosmic ray path length may be described in a LBM in a manner consistent with a mean path length, $\lambda = 20.6$ $\beta$ $P^{-0.45}$ at rigidities above ~0.5 GV (30 MeV/nuc for A/Z=2 nuclei). Both $^2$H and Boron have an excess intensity vs. the predictions for path lengths of 9 g/cm$^2$ below ~30 MeV/nuc.

**Introduction**

The spectra of H and He nuclei, along with their isotopes $^2$H and $^3$He, provide a unique perspective on the propagation of galactic cosmic rays in the galaxy. These 4 isotopes, with $^2$H and $^3$He as purely secondary nuclei and $^1$H and $^4$He as their main primary progenitors, form a "quartet" as it were. The study of these nuclei has had a long and somewhat controversial history following the initial studies and modelling by Comstock, et al., 1972. Earlier Voyager measurements of this quartet (Webber and Yushak, 1983; Seo, et al., 1994) and also more recent measurements at higher energies by PAMELA (Adriani, et al., 2015) near the Earth have greatly helped in trying to understand the characteristics of the local interstellar spectra of $^3$He and $^2$H nuclei. However, this earlier work did not prepare us for what has been actually observed at Voyager 1 at energies below ~100 MeV/nuc and described here.

One would be remiss in not noting the pioneering work on the production of the $^2$H and $^3$He nuclei during interstellar propagation by J.P. Meyer, 1972, and the more recent studies of this secondary production by Coste, et al., 2012, much of what is used in the cross section base for propagation calculations in this paper.

**The Data**

The $^2$H and $^3$He spectra as well as the primary $^1$H and $^4$He spectra measured at Voyager are shown in Table 1 and in Figures 1 and 2. They are obtained from the B end analysis of the HET-2 telescope on V1 (Stone, et al., 1977) during a 5 year time interval from the end of 2012 to the end of 2017 after V1 crossed the heliopause. An outline drawing of the telescope used here is shown in Figure 3. This telescope has been used to measure the spectra of the quartet in the two earlier publications noted in the introduction. In the bottom part of Figures 1 and 2 are some of these earlier measurement of these two secondary isotopes from V2 in 1977 and 1987 (Webber and Yushak, 1983; Seo, et al., 1994) and also measurements at higher energies on PAMELA in 2006-2007, (Adriani, et al., 2015). All of these earlier measurements were made at a typical solar modulation level of about ~400 MV in the inner heliosphere.

### a)  Mass Analysis – H Isotopes

For the B end of the analysis, a matrix of events (B1+B2) vs. C2-4 for H and $^2$H, subject to B1 vs. B2 consistency criteria, is first made.  This matrix for H isotopes is shown in Figure 4. The horizontal solid lines at various C2-4 pulse heights along the $^2$H mass line that is visible on the matrix define the energy loss intervals in C2-4 used in the determination of the $^2$H energy spectrum.  The mass line for $^2$H stands out clearly from the "background" between H and $^2$H which overall amounts to a 6-20% correction of the total number of $^2$H events in each energy interval, with this correction increasing with decreasing  energy (see also discussion by Seo, et al., 1994, who arrive at similar background corrections for $^2$H).

### b)  Mass Analysis – He Isotopes

Figure 5 shows a histogram of the number of events as a function of distance from the $^4$He mass line for $^3$He and $^4$He for the B end telescope.  The total background subtraction for events in the $^3$He mass distribution, which is assumed to be a Gaussian distribution for $^3$He and an exponential distribution for the $^4$He background, is shown by a dashed line in Figure 5 and this background is between 10-15% of the total unsubtracted numbers of events between the vertical dashed lines in Figure 5.  The background at a mass = 3.0 AMU is ~9% of the total events in the peak channel of the $^3$He distribution and the peak/valley ratio at the minimum between $^3$He and $^4$He is typically ~5 to 1.

### Summary of Mass Analysis

The results for the H and He intensities in the energy intervals for each isotope along with the estimated background corrections in percent for $^2$H and $^3$He for the B-end telescope is shown in Table 1.

The spectra in Figures 1 and 2, derived from the analysis which is summarized in Table 1, are what are needed to be fit by galactic propagation models for these secondary nuclei.  Most of the earlier measurements of the $^2$H and $^3$He secondaries were made between ~100 MeV/nuc and ~1 GeV/nuc and at locations where solar modulation effects were considerable as can be seen from the earlier data in Figures 1 and 2.  The earlier measured intensities of $^2$H and $^3$He nuclei below ~100 MeV/nuc lie a factor of nearly 10 below the recent Voyager measurements which are, in fact, the interstellar intensities. The new V1 data in Figures 1 and 2 may be compared with that from Seo, et al., 1994, from V2 when it was at 26 AU and from PAMELA (Adriani, et al., 2015) at higher energies near the Earth. The new interstellar spectra may also be compared with LBM galactic propagation calculations which are described in the following section.

## The Calculation of Secondary $^2$H and $^3$He Production in the Galaxy and a Comparison with Voyager Observations

This secondary production calculation is made with a LBM program that has been used in the past to calculate the local interstellar spectra of charges and isotopes from $^1$H to $^{62}$Ni (Webber and Higbie, 2009; Cummings, et al., 2016). In this calculation for the H and He isotopes the most important factors are:

(1) The mean path length in the galaxy, which is taken here to be $\lambda = 20.6 \beta P^{-0.45}$ g/cm$^2$ above P=0.5 GV rigidity in a LBM calculation (Webber, et al., 2017). The average matter density is 0.4 cm$^{-3}$, consisting of 90% H and 10% He, with 15% of the H ionized. This path length is largely determined above about 2.0 GV by a fit of the LBM propagation calculations of the B/C ratio to the observed secondary B/C ratio above a few hundred MeV/nuc (Webber and Villa, 2016, 2017) which includes earlier HEAO-3 data, (Engelmann, et al., 1990), and recent measurements of the B/C ratio by AMS-2 up to several hundred GeV/nuc (Aguilar, et al., 2016). These measurements have now defined the B/C ratio above a few 100 MeV/nuc to within $\pm$ a few percent. Below ~200 MeV/nuc the ACE measurements of this ratio, corrected for solar modulation may be also used (e.g., Lave, et al., 2013) in addition to the V1 measurements of the B/C ratio (Cummings, et al., 2016).

This higher rigidity interstellar path length, extended to lower rigidities, is shown in Figure 6. This path length above $P_0$=0.5 GV and with a path length $\sim\beta$ below $P_0$ provides a good fit in a LBM calculation to the individual H and He nuclei spectra between about 30 MeV/nuc and 350 MeV/nuc that are measured by V1 (Cummings, et al., 2016).

In the above expression the value of $\lambda$ reaches a maximum ~10 g/cm$^2$ at about 1.76 GV (400 MeV/nuc for A/Z = 2 nuclei). At high rigidities it decreases $\sim P^{-0.45}$ and at 200 GV (~100 GeV/nuc for A/Z = 2 nuclei) it is still ~1.6 g/cm$^2$.

At rigidities less than ~1.76 GV the path length decreases ~$\beta$ $P^{-0.45}$ to a value ~8.0 g/cm$^2$ at an energy of 30 MeV/nuc (for A/Z = 2 nuclei, $P_0$ ~0.562 GV) if the diffusion coefficient continues to be ~$P^{0.45}$ as it is at high rigidities. At still lower rigidities the effective mean path length is assumed to change to become ~$\beta$ below 0.5 GV. This low rigidity dependence has been determined by fitting the electron spectrum measured on V1. This electron spectrum is measured to be ~$E^{-1.35}$ between ~3-70 MeV whereas the source spectrum is believed to be ~$P^{-2.25}$. This one power difference in spectral index between the source spectrum and the spectrum measured at Voyager may be explained by a diffusion coefficient for electrons that is ~$P^{-1.0}$ at lower rigidities (see Webber and Villa, 2016, 2017). This simple form for the path length with a single break in the rigidity dependence at or below about 0.5 GV, gives spectra for H and He nuclei and also, very importantly, for the electrons just described, which are a good fit to the Voyager measurements of both of these types of particles between a few MeV/nuc and a few GeV/nuc (Webber and Higbie 2015; Webber and Villa, 2017).

This path length dependence is shown in Figure 6 for various other possible values of $P_0$. If the change in path length dependence occurs at $P_0$=1.0, for example, the mean path length is only about 6 g/cm$^2$ at 30 MeV/nuc (for A/Z=2.0 nuclei) as a result of the dependence of $\lambda$ ~$\beta$ below 1.0 GV. These low rigidity path lengths may be compared to the path length used by Lave, et al., 2013 also shown in Figure 6, which only extends down to~1 GV because of limitations due to solar modulation. This also approaches a $\lambda$ ~$\beta$ type of dependence at the lowest rigidities but implies much lower values for the path length.

(2) The cross sections for fragmentation into secondary nuclei from the primary or source nuclei. In this paper we are mainly interested in the fractions $^4$He into $^3$He and $^2$H. These interactions produce ~75% of all secondary $^3$He and $^2$H with the remainder being produced by C, O and Fe nuclei interactions (e.g., Coste, et al., 2012). In Table 2 we list these cross sections for $^4$He into $^3$He and $^2$H used in the LBM calculation. This includes all of the $^4$He into $^3$He cross sections, (including → $^3$He, $^3$H) and also (d+$^3$He) and the cross sections into $^2$H including again the (d+$^3$He) cross section and also the 2x$^2$H cross sections, the p+p reactions and various other $^2$H production channels. Again we stress the significance of the work of both Myer, 1972 and Coste, et al., 2012, in our choice of cross sections. Also of particular importance here are the stripping (elastic) cross sections (e.g., of the type $^4$He→$^3$He or $^3$H, $^{12}$C→$^{11}$C) which become large at energies <100 MeV/nuc. See Cumming, 1963, for $^{12}$C→$^{11}$C cross sections down to ~30 MeV/nuc. These cross sections are expected to be similar to the $^{4}$He→$^{3}$He, $^{3}$H cross section dependence at low energies. Many of these cross sections are underestimated or not even included in earlier studies, but are very important for the Voyager measurements.

The heavier nuclei cross sections, e.g., C, O, Fe→$^{2}$H and $^{3}$He are used in the propagation calculations but not shown here. The precision of all of these cross sections decreases considerably below ~100 MeV/nuc where the stripping cross sections are large and where the accuracy is probably $\pm$ 10-20 % at best.

The calculated secondary production of $^{2}$H and $^{3}$He is shown in Figures 1 and 2 for values of $P_0$ = 0.316, 0.562 and 1.0 GV with a dependence of $\lambda \sim \beta$ below each value of $P_0$ chosen. At a rigidity of $P_0$ the interstellar diffusion coefficient is assumed to change from a $P^{0.45}$ dependence at higher rigidities to a $P^{-1.0}$ dependence at lower rigidities thus producing the decreasing path lengths ~$\beta$ at lower energies as illustrated in Figure 6. This type of change in rigidity dependence of the diffusion coefficient at or below ~1 GV has been suggested earlier on theoretical grounds by Ptuskin, et al., 2006 (see also Alosio, Blasi and Serpico, 2016).

This secondary production is also shown for a constant path length = 9.0% g/cm$^2$ below $P_0$=1.0 GV as a possible upper limit to the interstellar production of $^{2}$H and $^{3}$He at lower energies.

## Comparison of Observations and Predictions

The $^{2}$H and $^{3}$He intensities measured at Voyager above ~20-30 MeV/nuc shown in Figures 1 and 2 generally are about equal to or slightly larger than the predictions of the LBM for values of $P_0$ between 0.316 and 0.562 GV. This excess over predictions increases with decreasing energy and is more prominent for $^{2}$H. This would mean that the mean path length traversed by cosmic rays as illustrated in Figure 6 is between 7-9 g/cm$^2$ at rigidities below 1.0 GV.

These measured $^{2}$H and $^{3}$He intensities are directly sensitive to the amount of material traversed by cosmic rays in the galaxy in the energy range 20-120 MeV/nuc corresponding to rigidities from ~0.25-1.00 GV.

Other than $^2$H and $^3$He the only other purely secondary nucleus with sufficient statistics from the CRS instrument on V1 to accurately determine the interstellar path length is Boron. The Boron spectrum also has been measured on V1 (Cummings, et al., 2016). In Figure 7 we show the spectra of $^2$H and $^3$He reported in this paper along with that for B, presented in Cummings, et al., 2016, side by side. For each spectrum we show the calculated results from our LBM for values of $P_0$=0.316 and 0.562 GV as well as the $\lambda$=const = 9.0 g/cm$^2$ below 1.0 GV example as described earlier. Above ~30 MeV/nuc all three nuclei are consistent with the assumption of a mean path length $\lambda$ ~$\beta$ P$^{-0.45}$ extending down to rigidities ~0.562 GV. This implies a matter path length traversed by primary cosmic ray He and heavier nuclei in a LBM that is ~7 g/cm$^2$ to ~9 g/cm$^2$ between 30 and 100 MeV/nuc. Below ~30 MeV/nuc all three measured intensities, $^2$H, $^3$He and B, exceed the predictions even for a mean path length 9.0 g/cm$^2$ of material.

In Figure 8 we show the Voyager observations of the $^3$He/$^4$He and $^2$H/$^4$He ratios along with the predictions for the LBM with the path length $\lambda$=20.6 $\beta$ P$^{-0.45}$ above 1.0 GV and equal to 9 g/cm$^2$ below 1.0 GV. These ratios provide an alternative perspective on the comparison of calculated and measured ratios. Again, above ~30 MeV/nuc the measured and calculated ratios agree.

## Summary and Conclusions

A 5 year period of data analysis at V1 has provided interstellar spectra of the secondary cosmic ray nuclei $^3$He and $^2$H of useful statistical accuracy in the energy range ~20 to 85 MeV/nuc. These spectra are capable of providing a definitive measure of the amount of interstellar matter their progenitor nuclei, in this case mainly $^4$He, has passed through after acceleration. It is found that the mean path length for both cosmic ray $^2$H and $^3$He production is at least 7-9 g/cm$^2$ at energies below ~100 MeV/nuc, assuming a LBM propagation. The largest errors are estimated to be the uncertainties in V1 data and the cross sections, both of which are at least $\pm$ 10-20% at these low energies.

Within the framework of a Leaky Box propagation model this implies that if the path length is indeed equivalent to $\lambda$ = 20.6 $\beta$P$^{-0.45}$ at higher energies, as determined from other studies made on Voyager near the Earth, e.g., the Boron abundance and the B/C ratio (Webber and Villa, 2016, 2017), as well as the H and He spectra themselves (Cummings, et al., 2016), then the same dependence of the path length, $\sim P^{-0.45}$, that is observed at higher energies continues down to lower energies ($\sim$30 MeV/nuc or 0.5 GV).

The data from all three of the most abundant purely secondary nuclei, $^2$H, $^3$He and B above $\sim$30 MeV/nuc and up to $\sim$100 MeV/nuc are consistent with the scenario described above. Below 30 MeV/nuc the Boron intensities described in Cummings, et al., 2016, exceed these predictions.

**Acknowledgements:**  The authors are grateful to the Voyager team that designed and built the CRS experiment with the hope that one day it would measure the galactic spectra of nuclei and electrons.  This includes the present team with Ed Stone as PI, Alan Cummings, Nand Lal and Bryant Heikkila, and to others who are no longer members of the team, F.B. McDonald and R.E. Vogt.  Their prescience will not be forgotten.  This work has been supported throughout the more than 40 years since the launch of the Voyagers by the JPL.

| Emin | Emax | Total Cnts*/(% B) | Cnts/MeV/nuc | nact factor | Livetime | Geom Fact | Intensity + |
|---|---|---|---|---|---|---|---|
| **TABLE I** <br> **B STOP HIGH GAIN** <br> **2012.9-2017.75** | | | | | | | |
| **¹H** | | | | | | | |
| 28.54 | 30.5 | 26,500 | 13,520 | x1.005 | 2.766E+06 | 1.69E-04 | 29.1 |
| 30.5 | 36.75 | 83,000 | 13,200 | x1.007 | 2.766E+06 | 1.63 E-04 | 29.5/12.61 |
| 36.75 | 46.9 | 116,800 | 11,500 | x1.012 | 2.766E+06 | 1.46 E-04 | 28.8/12.42 |
| 46.9 | 59.9 | 123,000 | 9,460 | x1.018 | 2.766E+06 | 1.26 E-04 | 27.7/12.76 |
| 59.9 | 73.6 | 103,500 | 7,554 | x1.024 | 2.766E+06 | 1.05 E-04 | 26.8/12.76 |
| **²H** | | | | | | | |
| 20.7 | 25.0 | 501  -  (22%) | 116.5 | x1.01 | 2.766E+06 | 1.63 E-04 | 0.261 |
| 25.0 | 31.9 | 698  -  (20%) | 103.5 | x1.02 | 2.766E+06 | 1.46 E-04 | 0.261 |
| 31.9 | 40.7 | 726  -  (15%) | 82.6 | x1.03 | 2.766E+06 | 1.26 E-04 | 0.244 |
| 40.7 | 50.0 | 646  -  (6%) | 69.5 | x1.04 | 2.766E+06 | 1.05 E-04 | 0.250 |
| **⁴He** | | | | | | | |
| 30.5 | 36.75 | 6,488 | 1,038 | x1.020 | 2.766E+06 | 1.63 E-04 | 2.35 |
| 36.75 | 46.9 | 9,282 | 915 | x1.025 | 2.766E+06 | 1.46 E-04 | 2.32 |
| 46.9 | 59.9 | 9,494 | 730 | x1.035 | 2.766E+06 | 1.26 E-04 | 2.17 |
| 59.9 | 73.6 | 7,940 | 580 | x1.05 | 2.766E+06 | 1.05 E-04 | 2.10 |
| **³He** | | | | | | | |
| 35.9 | 43.3 | 487  -  (13%) | 65.6 | x1.018 | 2.766E+06 | 1.63 E-04 | 0.147 |
| 43.3 | 55.4 | 754  -  (13%) | 62.1 | x1.02 | 2.766E+06 | 1.46 E-04 | 0.157 |
| 55.4 | 70.5 | 841  -  (13%) | 55.7 | x1.03 | 2.766E+06 | 1.26 E-04 | 0.165 |
| 70.5 | 86.7 | 771  -  (12%) | 48.2 | x1.04 | 2.766E+06 | 1.05 E-04 | 0.173 |

\* Total counts for ³He are the numbers of events between masses of 2.5 and 3.5 with a percent background subtracted as indicated in column 3.

\+ All intensities are in Particles/m²·sr·s·MeV/nuc $\pm$ 5%. For the H intensities the number after the slash refers to the total H over the total He ratio.

| TABLE II Cross Sections in mb | | | | | | | | | | | | | | |
|---|---|---|---|---|---|---|---|---|---|---|---|---|---|---|
| **Energy (MeV/nuc)** | **10** | **20** | **40** | **60** | **80** | **100** | **150** | **200** | **300** | **400** | **600** | **800** | **1000** | **2000** | **5000** |
| He4 Total | 104 | 112 | 112 | 104 | 100 | 90 | 76 | 68 | 68 | 68 | 76 | 88 | 96 | 100 | 104 |
| He4→He3 | 48 | 100 | 100 | 92 | 80 | 72 | 65 | 60 | 56 | 54 | 53 | 53 | 54 | 54 | 54 |
| He4→H2 | 120 | 142 | 120 | 96 | 82 | 72 | 66 | 57 | 56 | 56 | 56 | 56 | 56 | 55 | 54 |
| He3 Total | 96 | 102 | 104 | 100 | 84 | 76 | 70 | 62 | 62 | 68 | 66 | 72 | 84 | 88 | 88 |
| H2 Total | 76 | 76 | 76 | 67 | 60 | 55 | 52 | 51 | 50 | 50 | 52 | 56 | 60 | 62 | 63 |

The individual cross sections to $^3$He and $^2$H are the total of all processes.     For $^3$He they include p+$^4$He→$^3$He, $^3$H and d+$^3$He

For $^2$H they include p+$^4$He→$^2$H, 2x$^2$H, d+$^3$He and

p+p→$^2$H ($\equiv$d)

**FIGURE CAPTIONS**

**Figure 1:** $^2$H and $^1$H abundance measured at Voyager for B stop nuclei. Calculated values of secondary production in a LBM for galactic propagation with $\lambda$=20.6 $\beta$ $P^{-0.45}$ above break rigidities $P_0$ of 0.316, 0.562 and 1.0 GV are shown along with a mean path length = 9.0 g/cm$^2$ below 1.0 GV. The solar modulation for a modulation potential of 400 MV is shown as a dashed line. Measurements of $^2$H at V1 in 1977 (Webber and Yushak, 1983), and 1987 (Seo, et al., 1994), and PAMELA in 2006-07 (Adriani, et al., 2015) are also shown.

**Figure 2:** Same as Figure 1 except $^3$He and $^4$He measurements and calculations are shown.

**Figure 3:** Outline drawing of HET telescope on the CRS experiment used to measure the spectra of H and He nuclei (see description by Stone, et al., 1977).

**Figure 4:** Matrix of events for B stop in C2-4, for $^1$H and $^2$H nuclei. A clearly defined $^2$H mass line is evident in this background subtracted image. The energy spectra of $^2$H and $^1$H are derived from the E loss distribution of events in the 4.5 g/cm$^2$ thick total energy counter, C2-4 (which is a composite of 3 counters, each 1.5 g/cm$^2$ thick), between the boundary lines at a constant value of C2-4 (energy loss) shown for $^2$H in this figure.

**Figure 5:** Mass histogram for $^3$He and $^4$He nuclei for B stop in C2-4. Dashed line shows the estimated background subtraction. Vertical dashed lines define the reference $^3$He distribution.

**Figure 6:** Mean path length as a function of rigidity for $\lambda$ = 20.6 $\beta P^{-0.45}$ above $P_0$ = 1.0 GV ($\sim$124 MeV/nuc for A/Z = 2.0 nuclei) and for several values of $P_0 < 1.0$ GV. Below $P_0$ the path length, $\lambda$, is taken to be $\sim\beta$. The energy/nuc for A/Z = 2.0 particles is also shown on the X-axis. The path length used for GALPROP calculations above $\sim$1 GV by Lave, et al., 2013, is shown as a red line.

**Figure 7:**  A comparison of the measurements of $^2$H, $^3$He and B nuclei intensities measured by the CRS experiment on V1 and the predictions of the LBM for values of $P_0$=0.316, 0.562 and 1.0 GV and a constant value = 9.0 g/cm$^2$ below 1.0 GV as described in the text.  The energies for $^3$He and B are offset in energy by factors of 10 and 100 respectively.

**Figure 8:**  The observed and calculated LIS abundance ratios $^2$H/$^4$He and $^3$He/$^4$He for a LBM in which $\lambda$=20.6 $\beta$ $P^{-0.45}$ for P > 1.0 GV and $\lambda$= 9g/cm$^2$ for P < 1.0 GV.

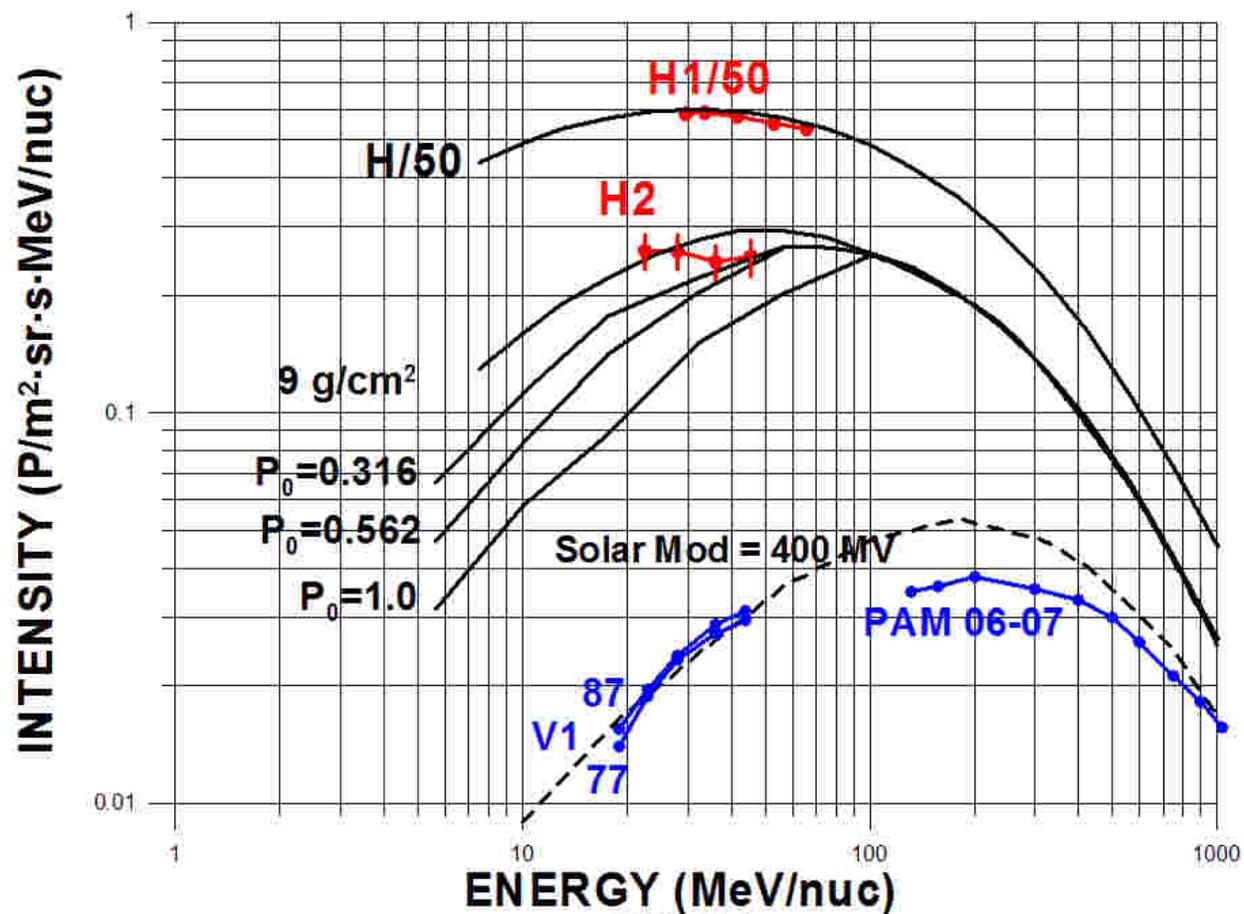

**FIGURE 1**

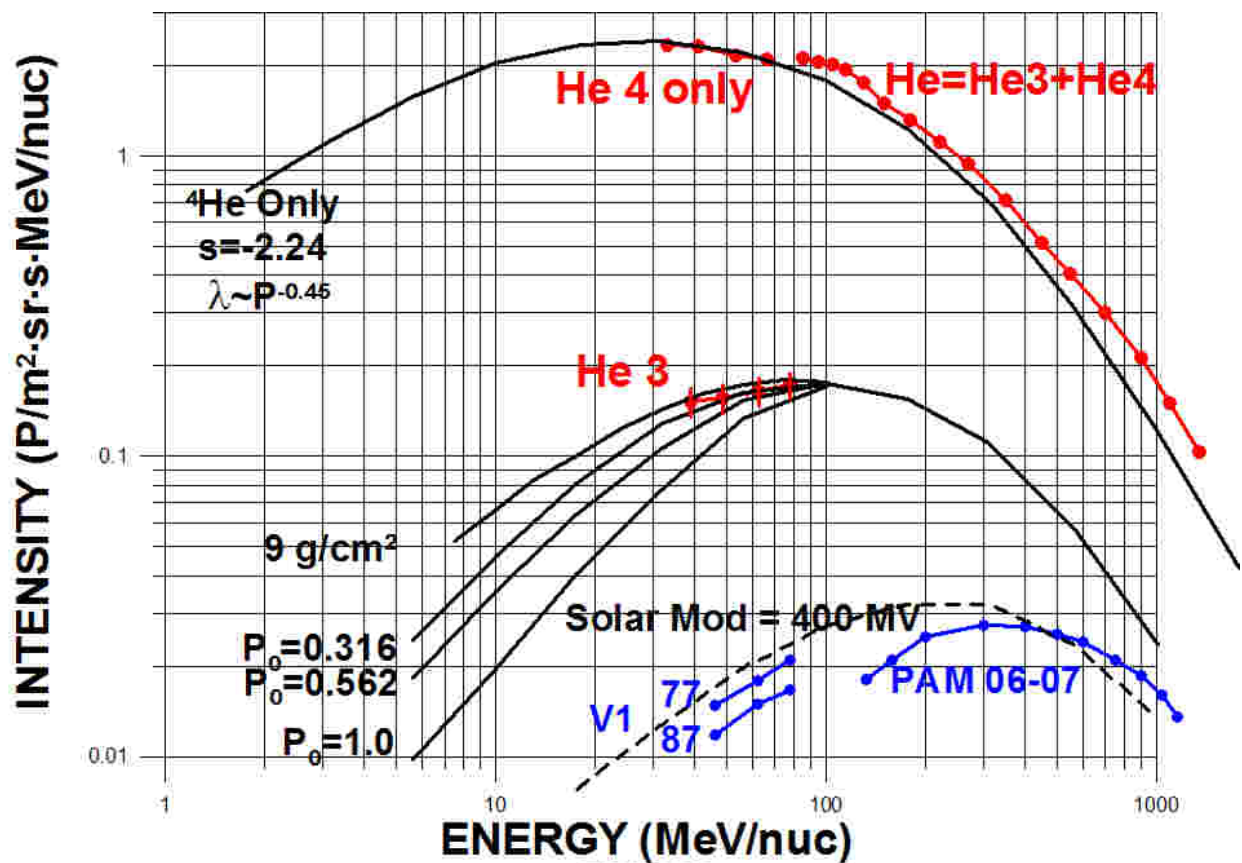

FIGURE 2

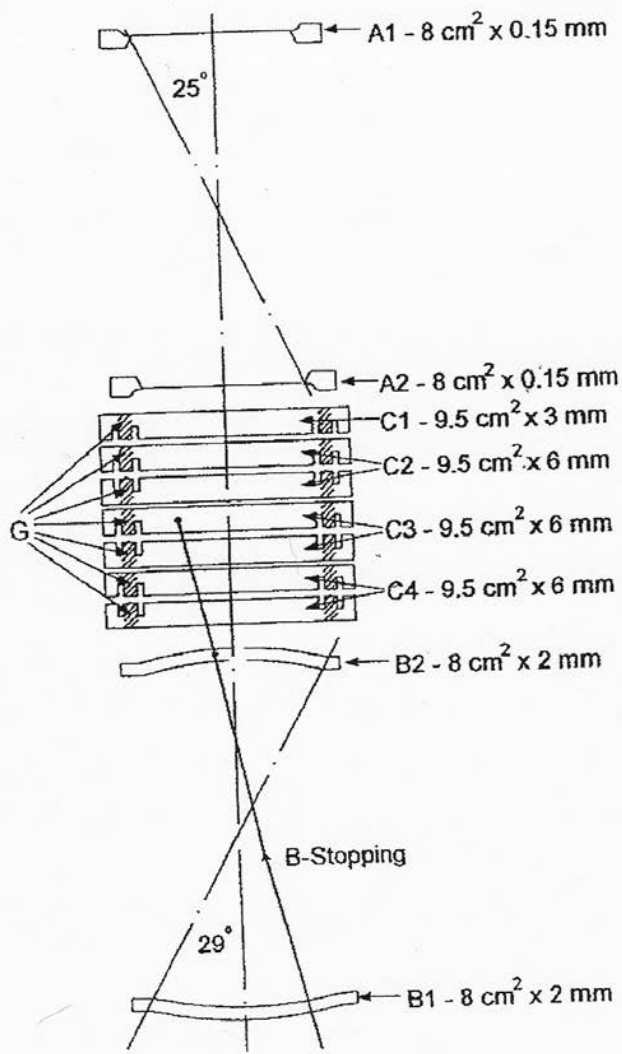

FIGURE 3

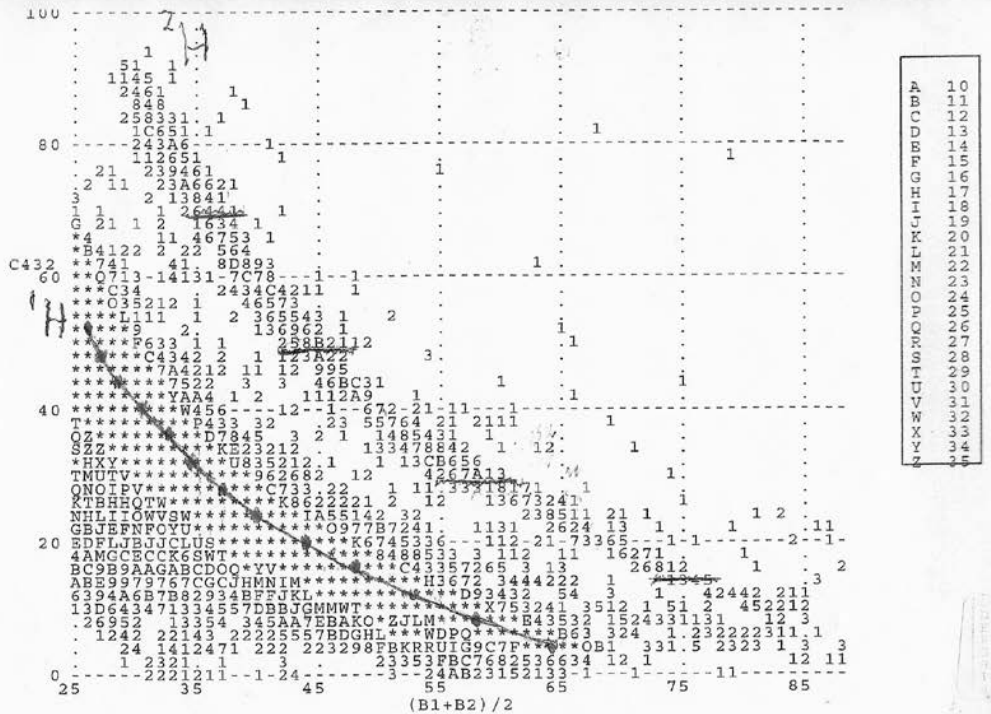

FIGURE 4

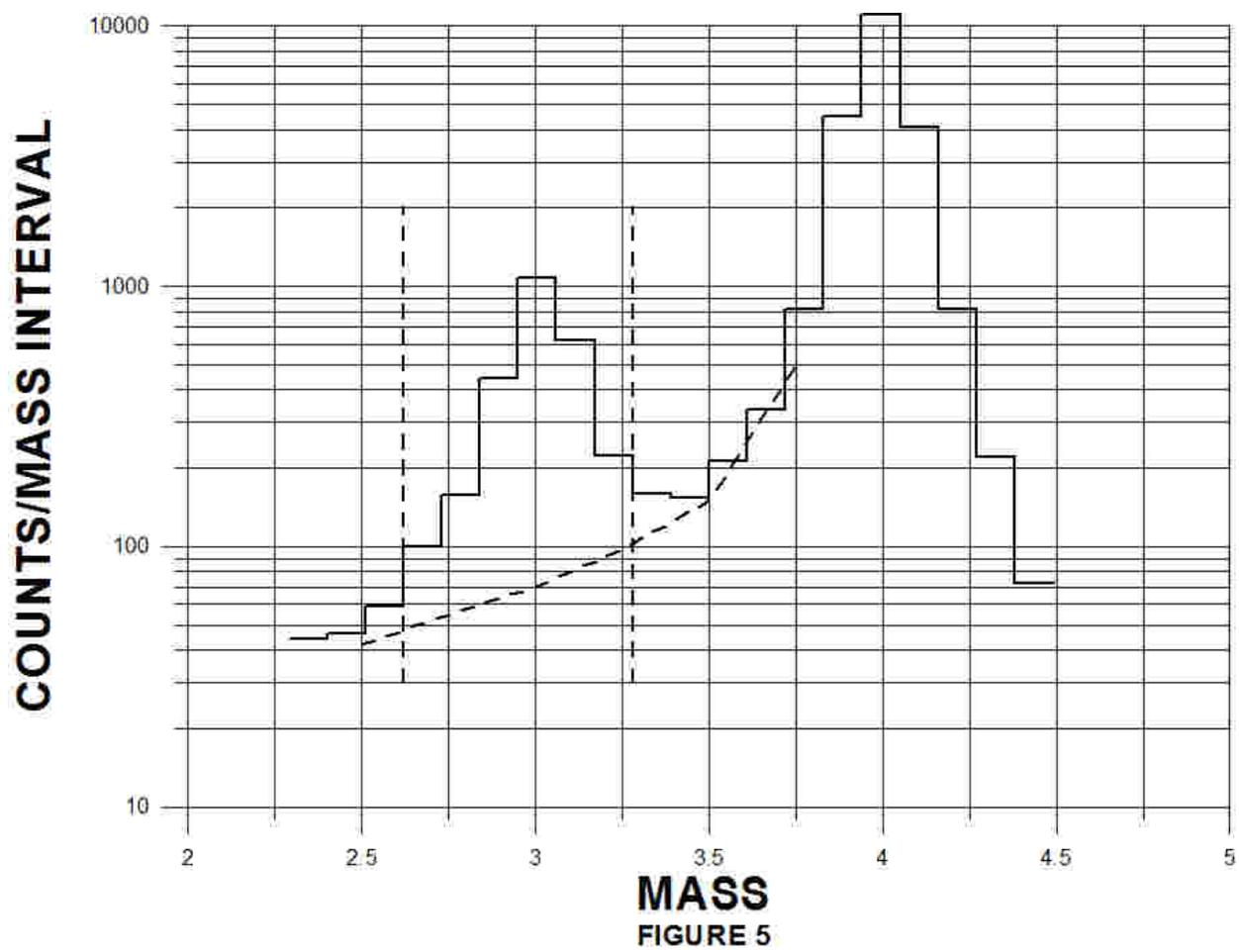

**MASS**

**FIGURE 5**

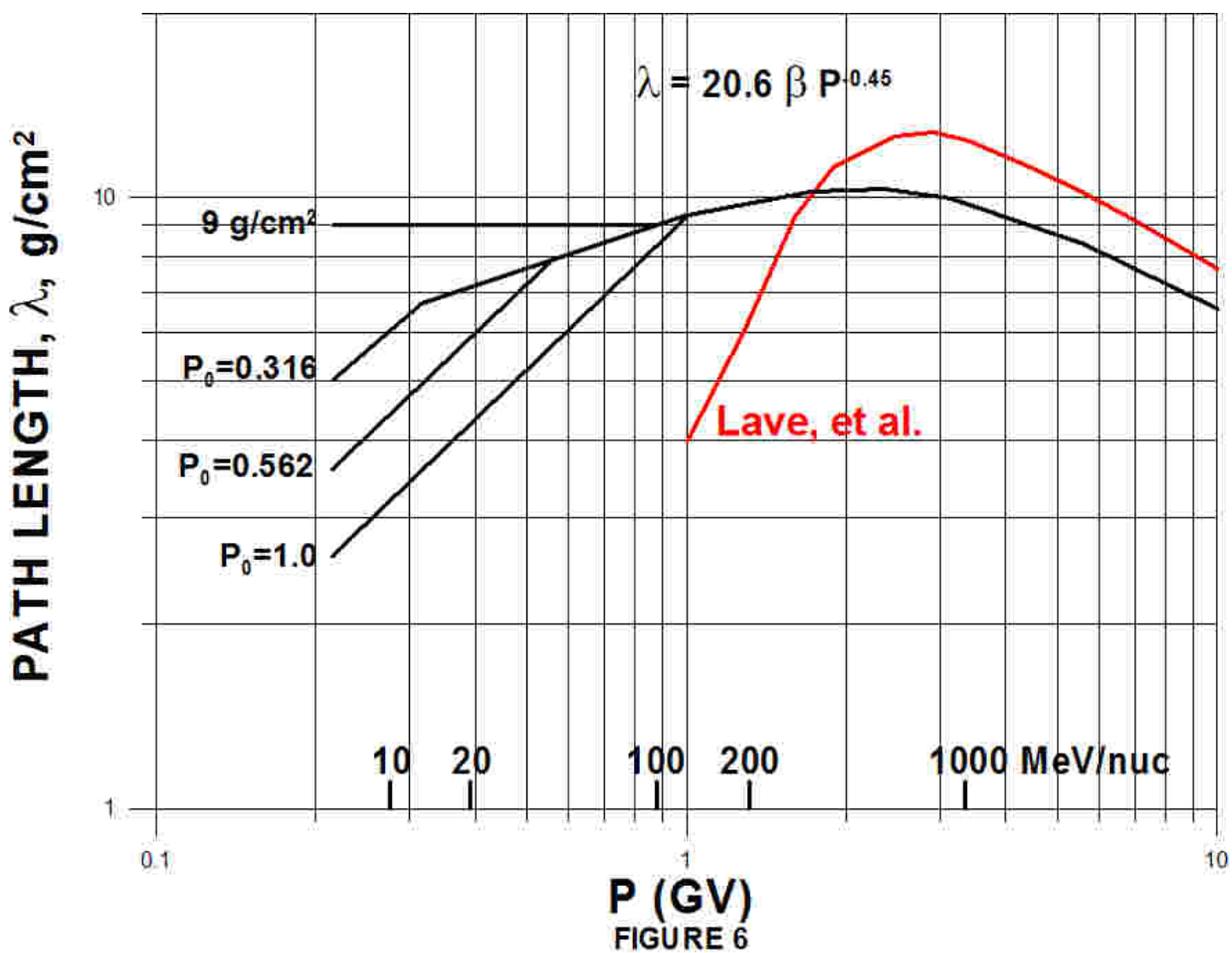

FIGURE 6

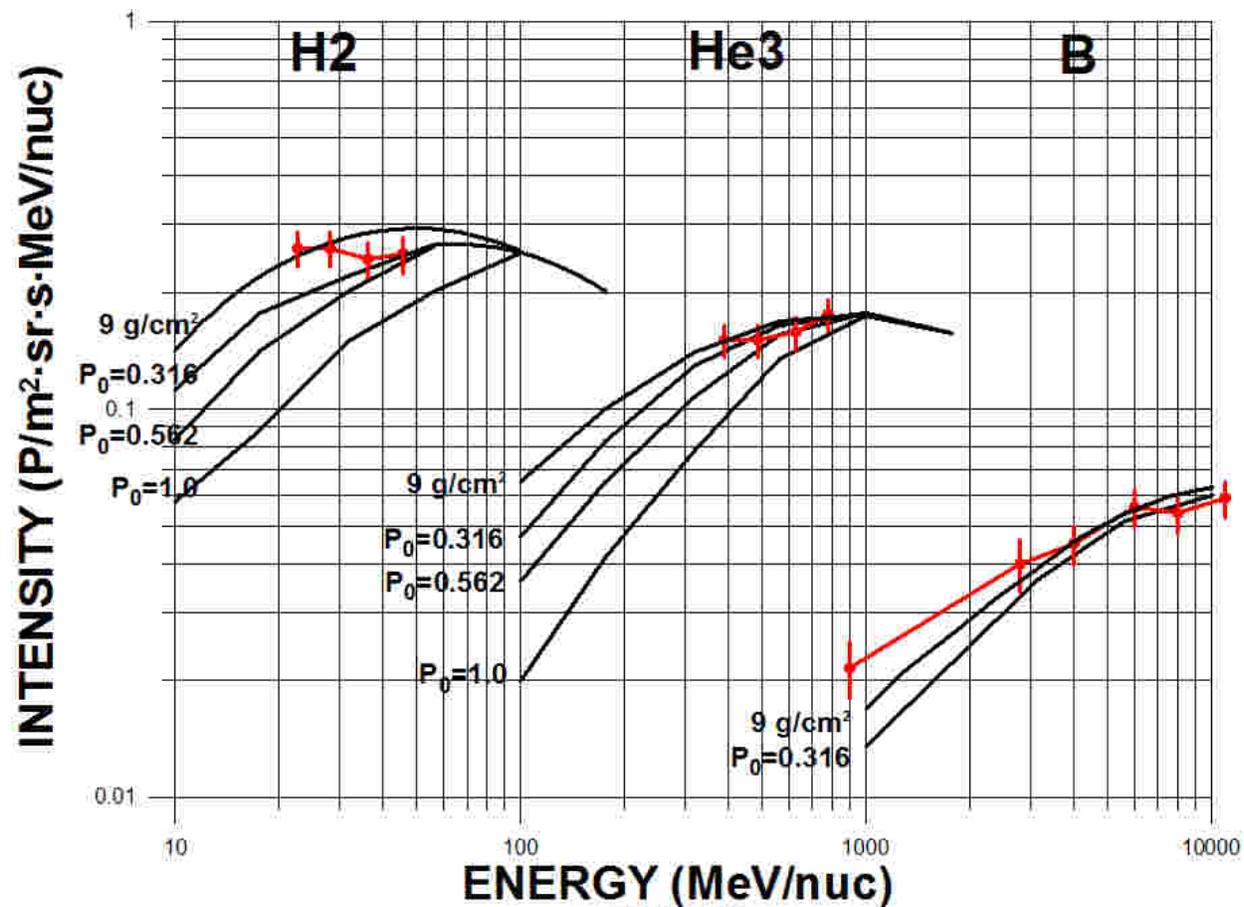

FIGURE 7

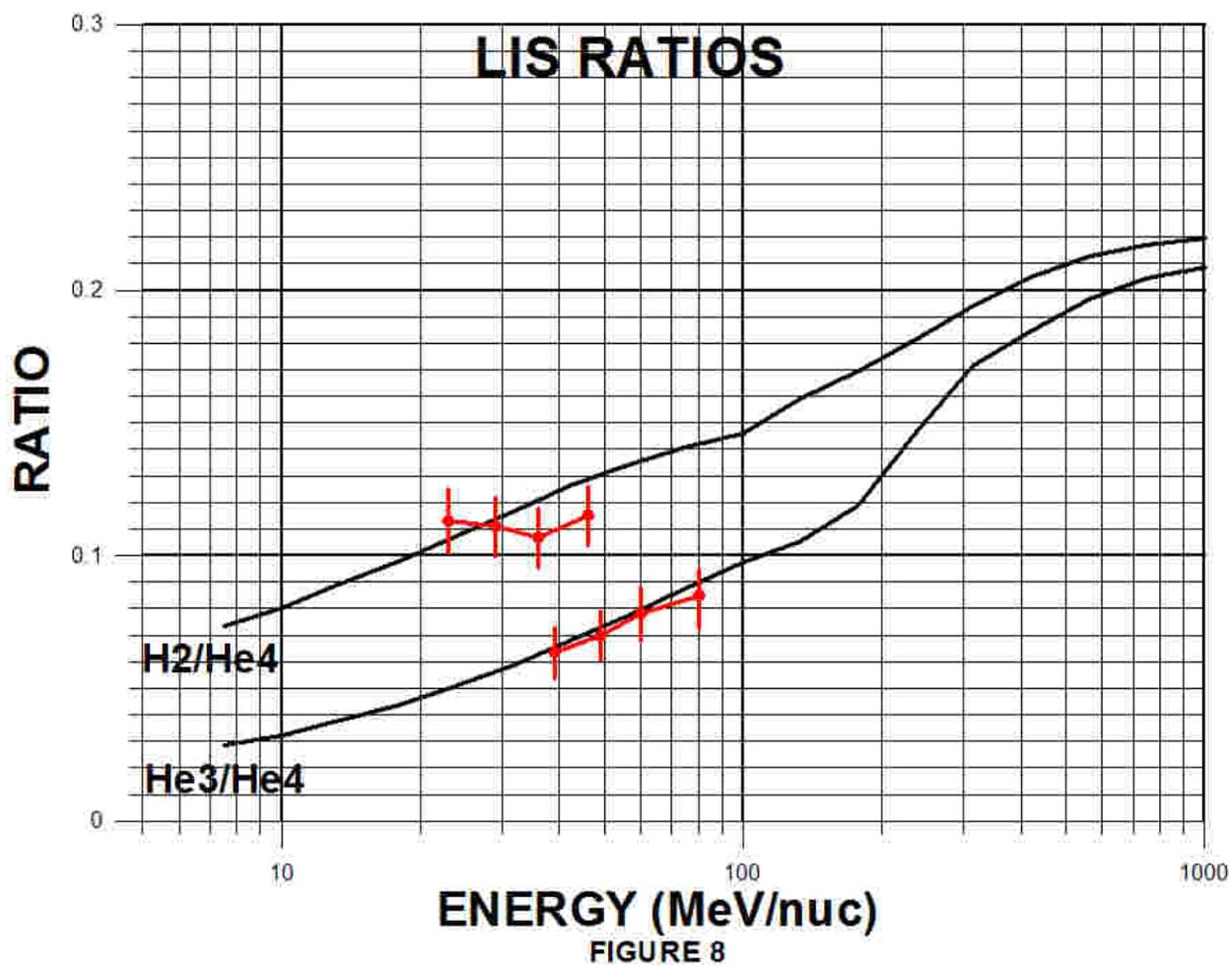

FIGURE 8